\documentclass[12pt]{article}
\usepackage{amsmath}
\usepackage{amstext}
\usepackage{amsfonts}
\usepackage{amsbsy}
\usepackage{eucal}
\usepackage{amssymb}
\usepackage{amsthm}
\usepackage{color}
\usepackage[dvips]{graphicx}

\theoremstyle{definition}

\begin{document}
\begin{center}
{\Large  Bilinear approach to the supersymmetric Gardner equation}
\end{center}
\begin{center}
N. C. Babalic, A. S. Carstea
\end{center}
\begin{center}
{\it Dept. of Theoretical Physics, Institute of Physics and Nuclear Engineering, P.O. BOX MG-6, Magurele, Bucharest, Romania }
\end{center}

\begin{abstract}
We study a supersymmetric version of the Gardner equation (both focusing and defocusing) using the superbilinear formalism. This equation is new and cannot be obtained from supersymmetric modified Korteweg-de Vries equation with a nonzero boundary condition. We construct supersymmetric solitons and then by passing to the long-wave limit in the focusing case obtain rational nonsingular solutions. We also discuss the supersymmetric version of the defocusing equation and the dynamics of its solutions.
\end{abstract}

\section{Introduction}
Studies of supersoliton equations have recently revealed very interesting aspects of supersoliton dynamics.
This progress was possible because of the extension of the Hirota bilinear formalism to superspace, which allowed computating a multisupersoliton solution for various equations 
\cite{fane1}-\cite{zhang1}. In contrast to the ordinary case, the interaction term 
contains a fermionic tail dressing the nonlinear mode.  Correspondingly, the more supersolitons there are involved in the interaction, the more ``dressed'' the interaction is in a specific way. 
It is especially interesting that the dressing has the same expression for many examples studied so far (KdV, mKdV, sine-Gordon, Sawada-Kotera, Ito, Calogero-Degasperis-Fokas). In the framework of the discrete 
setting, the dressing differs, but only the discretisation of supersymmetric KdV in the case of the supersymmetric KdV equation has been studied until now \cite{fane}.

Here, we apply the supersymmetric bilinear formalism to a new equation combining the supersymmetric KdV and mKdV equations. It is well known that in the classical context, the combined KdV+mKdV equation is usually called the Gardner equation and is essentially the mKdV equation with a nonzero constant boundary condition at infinity. This is not so in the supersymmetric case, and the combined super-KdV+super-mKdV equation is a new equation, which we call the super-Gardner equation. Applying the supersymmetric bilinear formalism, we show that it admits at least a three-supersoliton solution. 

In addition, we discuss the defocusing case and also the special solutions such as superrational and supershock solutions. In one-dimensional context, rational solutions are not very interesting, but the Gardner equation has nonsingular solutions with a Lorentzian shaped \cite{ablowitz}. In the supersymmetric context, we have an analogous situation, i.e., both the odd and even parts of the solution  have a nonsingular Lorentzian shape. Moreover, they have free parameters that cannot be gauged away (as in the ordinary case). We analyze the interaction of these superrational solutions with the supersolitons and show that the situation is analogous to the classical case: the supersoliton remains unchanged, and the superrational olution is shifted, in a certain sense resembling the dressing of the supersoliton interaction.

\section{Focusing super-Gardner equation}
We work in the $N=1$ superspace $(x,t,\theta)$, with only one independent fermionic variable $\theta$. We consider the equation:
\begin{equation}\label{s-gard}
\Phi_t+\Phi_{xxx}+3(D\Phi)(\Phi D\Phi)_x+3\sigma(\Phi D\Phi)_x=0,
\end{equation}
where $\Phi(x,t,\theta)=\zeta(x,t)+\theta u(x,t)$ is a fermionic superfield, $D=\partial_\theta+\theta\partial_x$ is the super-derivative, $\zeta(x,t)$ and $u(x,t)$ are fermionic and bosonic functions, 
and $\sigma$ is a bosonic arbitrary constant. Equation (\ref{s-gard}) is precisely the combined super-mKdV+super-KdV equation. 
We write it in components:
\begin{eqnarray*}
&&\zeta_t+\zeta_{xxx}+3 (u+\sigma) (\zeta u)_x=0\\
&&u_t+u_{xxx}+6 u^2 u_x+6\sigma u u_x-3u_x\zeta \zeta_x -3 (u+\sigma) \zeta \zeta_{xx}=0.
\end{eqnarray*}
If $\zeta(x,t) \rightarrow 0$, then we obtain the ordinary Gardner equation:
\begin{equation}\label{gard}
u_t+u_{xxx}+6 u^2 u_x+6\sigma u u_x=0.
\end{equation}
Generally speaking, the ordinary Gardner equation is the mKdV equation with a nonzero boundary condition. Nevertheless, Eq. (\ref{s-gard}) cannot be obtained from the supersymmetric mKdV equation by imposing a nonzero boundary $\sigma$. 
Indeed, it can be seen immediately that if we take the supersymmetric mKdV equation
\begin{equation}\label{mKdV}
\Psi_t+\Psi_{xxx}+3(D\Psi)(\Psi D\Psi)_x=0
\end{equation}
and set $\Psi=\phi+\beta$, where $\beta$ is an odd constant ( in the sense that it is independend of $x$ and $t$ but can depend on $\theta$, in which case $D\beta=\sigma$), 
then there is no way to obtain Eq. (\ref{s-gard}) for $\phi$  from (\ref{mKdV}). Consequently, Eq. (\ref{s-gard}) is new. 

To obtain the superbilinear form, we transform (\ref{s-gard}) into the potential form. Setting $\Phi=D\phi$, where $\phi(x,t,\theta)$ is a bosonic superfield, we write (\ref{s-gard}) as:
\begin{equation}\label{pot-gard}
\phi_t+\phi_{xxx}+2\phi_x^3-3(D\phi)(D\phi_x)\phi_x+3\sigma\phi_x^2-3\sigma D\phi D\phi_x=0.
\end{equation}
For superbilinearization, we take $\phi=i\log(g/f)$ with the bosonic superfunctions
$$g(x,t,\theta)=G+\theta \gamma, \hspace{1cm} f(x,t,\theta)=F+\theta \varphi.$$
Here, $F(x,t)$, $G(x,t)$ are bosonic (even) functions, and $\varphi(x,t)$ and$\gamma(x,t)$ are fermionic (odd) functions. 
Using the properties of the Hirota operator $\bf D$, we obtain
\begin{eqnarray}
\phi_t+\phi_{xxx}+2\phi_x^3&=&i\frac{({\bf D}_t+{\bf D}_x^3)g\cdot f}{g f}-3i\frac{({\bf D}_x^2g\cdot f)({\bf D}_xg\cdot f)}{g^2 f^2}, \nonumber \\ 
-3(D\phi)(D\phi_x)\phi_x&=&3i\frac{D(gf)({\bf S}_x^3g\cdot f)}{g^2 f^2}\frac{{\bf D}_x g\cdot f}{gf}+3i\frac{({\bf D}_x^2g\cdot f)({\bf D}_xg\cdot f)}{g^2 f^2}- \nonumber\\
&&- 3i\frac{D({\bf S}_x^3g\cdot f)({\bf D}_xg\cdot f)}{g^2 f^2}, \nonumber\\
3\sigma\phi_x^2&=&-3\sigma\left(\frac{{\bf D}_x g\cdot f}{gf}  \right)^2, \nonumber\\
-3\sigma D\phi D\phi_x&=&3\sigma\frac{D(gf)({\bf S}_x^3g\cdot f)}{g^2 f^2}+3\sigma\frac{{\bf D}_x^2 g\cdot f}{gf}-3\sigma\frac{D({\bf S}_x^3g\cdot f)}{gf}, \nonumber
\end{eqnarray}
where $S^N_x a \cdot b$ is the super-Hirota operator defined in \cite{fane1}, acting on two arbitrary superfunctions $a(x,\theta)$ and $b(x,\theta)$, and is given by:
$${\bf S}_x^{2N}a\cdot b={\bf D}^Na\cdot b,\quad {\bf S}_x^{2N+1}a\cdot b={\bf S}_x{\bf D}^Na\cdot b,$$
$${\bf S}_x a\cdot b=(Da)b-(-1)^{|a|}a(Db),$$
where $|a|$ is the grassmann parity of the superfunction $a(x,\theta)$.

Adding the above expressions together and collecting purely bilinear terms, we obtain:
\begin{equation}\label{dis}
 (i{\bf D}_t+i{\bf D}_x^3+3\sigma{\bf D}_x^2+{\rm possible\hspace{0.2cm}other\hspace{0.2cm}terms})g\cdot f=0.
\end{equation}
The remaining part is:
\begin{equation}\label{cons}
 \left(D(gf)({\bf S}_x^3g\cdot f)-gfD({\bf S}_x^3g\cdot f)\right)\left(3i{\bf D}_xg\cdot f+3\sigma gf\right)=3\sigma g f({\bf D}_xg\cdot f)^2.
\end{equation}
We also have the relation
\begin{equation}\label{rel}
 gf({\bf D}_xg\cdot f)D({\bf S}_xg\cdot f)-D(gf)({\bf S}_xg\cdot f)({\bf D}_xg\cdot f)=gf({\bf D}_xg\cdot f)^2,
\end{equation}
which can be verified by direct calculation.
Replacing the right-hand side of (\ref{cons}) with the left-hand side of (\ref{rel}) multiplied with $3 \sigma$, we obtain:
$$3 i D(gf)({\bf D}_x g\cdot f)({\bf S}_x^3 g\cdot f-i\sigma{\bf S}_xg\cdot f)-$$
$$-3igf({\bf D}_x g\cdot f)(D({\bf S}_x^3 g\cdot f-i\sigma{\bf S}_xg\cdot f))+$$
\begin{equation}\label{cons0}
+3\sigma g f D(gf)({\bf S}_x^3 g\cdot f)-3\sigma g^2f^2 D({\bf S}_x^3 g\cdot f)=0.
\end{equation}
From this expression, we impose the bilinear relation:
\begin{equation}\label{cons1}
{\bf S}_x^3 g\cdot f-i\sigma{\bf S}_xg\cdot f=0,
\end{equation}
which annihilates two terms in the first line in (\ref{cons0}). Replacing ${\bf S}_x^3$ with  $i\sigma{\bf S}_x$ in the remaining part of (\ref{cons0}), we obtain:
\begin{equation}\label{last}
3i\sigma^2 (g f D(gf)({\bf S}_x g\cdot f)-g^2f^2D({\bf S}_x g\cdot f))=0.
\end{equation}
We can immediately see that (\ref{last}) is just 
$$-3i\sigma^2g^2f^2({\bf D}_xg\cdot f),$$
which can be transferred to (\ref{dis}), replacing ``possible other terms''. Finally, the Hirota bilinear form of the focusing super-Gardner equation is written as:
\begin{eqnarray}\label{bilinear g}
({\bf D}_t+{\bf D}_x^3-3i\sigma{\bf D}_x^2-3\sigma^2{\bf D}_x)g\cdot f=0\\
({\bf S}_x^3 -i\sigma{\bf S}_x)g\cdot f=0. \nonumber
\end{eqnarray}

Using the simple linear transformation: $$\left\{ \begin{array}{rcl}
 x &\rightarrow& X-3 \sigma^2 T\\ t &\rightarrow& T \end{array}\right. ,$$ our system becomes simpler:
\begin{eqnarray}\label{bilinearr}
({\bf D}_T+{\bf D}_X^3-3i\sigma{\bf D}_X^2)g\cdot f=0\\
({\bf S}_X^3 -i\sigma{\bf S}_X)g\cdot f=0. \nonumber
\end{eqnarray}

Using the components $ g=G+\theta \gamma$ and $f=F+\theta \varphi$, where $G(x,t)$ and $F(x,t)$ are bosonic (even) fields and $\zeta(x,t)$ and $\phi(x,t)$ are fermionic (odd) fields, we write the system as:
\begin{eqnarray} \label{pe componente1}
({\bf D}_T+{\bf D}_X^3-3\sigma i {\bf D}_X^2) G \cdot F=0 \nonumber \\
({\bf D}_T+{\bf D}_X^3-3\sigma i {\bf D}_X^2) (\gamma \cdot F+ G \cdot \varphi)=0 \nonumber \\ 
({\bf D}_X-i \sigma)\gamma \cdot F - ({\bf D}_X-i \sigma)G \cdot \varphi=0 \nonumber \\
{\bf D}_X ({\bf D}_X -i \sigma)G\cdot F-2({\bf D}_X-i \sigma)\gamma \cdot \varphi=0. \nonumber
\end{eqnarray}

It can be seen that if $\gamma=0$ and $\phi=0$, then this system becomes classical bilinear form \cite{ablowitz} of Gardner equation (\ref{gard}):
\begin{eqnarray}
({\bf D}_T+{\bf D}_X^3-3\sigma i {\bf D}_X^2) G \cdot F=0 \nonumber \\
{\bf D}_X ({\bf D}_X -i \sigma)G\cdot F=0. \nonumber
\end{eqnarray}

{\bf Remark:} The bilinear system (\ref{bilinear g}) appeared in literature (see \cite{liu2}) as a superbilinear B\"acklund tranformation relating various solutions of the supersymmetric KdV equation. 

\section{Supersoliton solutions}
Soliton solutions of (\ref{bilinearr}) can be constructed using combinations of modes defined on the superspace $e^{\eta+\theta \xi}=e^{k X+\omega T+\eta_0+\theta \xi}$ (here $\eta_0$ is an 
arbitrary even constant that is crucial for computing rational solutions). Basically, any supersoliton is characterised by an even-odd pair of 
parameters $(k,\xi)$ corresponding to the coordinates of the superspace $(X,\theta)$ and  $\omega=\omega(k,\xi)$, generally speaking. 
Of course, nobody requires us to fix the number of odd parameters, and a supersoliton can therefore be characterised by many odd parameters like $(k,\xi_1,...\xi_p)$. But we want to construct the 
supersolitons as fundamental blocks 
depending on a minimum number of parameters, and we therefore consider only one pair $(k,\xi)$ for each supersoliton in what follows.
 
The tau functions $g$ and $f$ for the one-supersoliton have the forms:
\begin{eqnarray}\label{1sol}
f&=&1+a e^{\eta+\theta\xi},\\
g&=&1+a^* e^{\eta+\theta\xi}= f^*,\nonumber
\end{eqnarray}
where $\eta=k X+\omega T+\eta_0$, $\omega(k)=-k^3-3 k \sigma^2$, $a=1 + i \frac{k}{\sigma}$ and $\xi$ is an arbitrary Grassmann-odd constant. The one-supersoliton has the form:
\begin{eqnarray}\label{1 ss susyGardner}
\Phi_{1ss}&=&D \phi=iD\log \frac{g}{f}=iD \log \frac{1+a^* e^{\eta+\theta\xi}}{1+ae^{\eta+\theta\xi}}=\\
&=&2 \frac{k}{\sigma} \frac{e^{\eta}}{\left(1+a^* e^{\eta}\right)\left(1+a e^{\eta}\right)}\xi-2 \frac{k^2}{\sigma} \frac{e^{\eta}}{\left(1+a^* e^{\eta}\right)\left(1+a e^{\eta}\right)}\theta= \nonumber\\
&=&\frac{k}{\sigma}\frac{\xi}{1+\sqrt{a a^*}\cosh \left(\eta+\log{\sqrt{a a^*}}\right)}+ \frac{k^2}{\sigma}\frac{\theta}{1+\sqrt{a a^*}\cosh \left(\eta+\log{\sqrt{a a^*}}\right)}. \nonumber
\end{eqnarray}

For the two-supersoliton solution, we now have two pairs of parameters $(k_1,\xi_1)$ and $(k_2,\xi_2)$ for each supersoliton, and the interaction term must be added, of course.  We here give only the expressions for tau functions $g$ and $f$ because the nonlinear form is complicated:
\begin{eqnarray}\label{2 ss Gardner}
f=1 +a_1 e^{ \eta_1+\theta\xi_1}+a_2e^{\eta_2+\theta\xi_2}+ A_{12} a_1 a_2(1+\beta_{12}\xi_1\xi_2)e^{ \eta_1 + \eta_2+\theta(\alpha_{12}\xi_1+\alpha_{21}\xi_2)},\\ \nonumber
g=1 +a^*_1 e^{ \eta_1+\theta\xi_1}+a^*_2e^{\eta_2+\theta\xi_2}+ A_{12} a^*_1 a^*_2(1+\beta_{12}\xi_1\xi_2)e^{ \eta_1 + \eta_2+\theta(\alpha_{12}\xi_1+\alpha_{21}\xi_2)},
\end{eqnarray}
where
$$A_{ij}=\left(\frac{k_i-k_j}{k_i+k_j}\right)^2,  \quad \beta_{ij}=\frac{2}{k_i-k_j},  \quad \alpha_{ij}=-\alpha_{ji}=\frac{k_i+k_j}{k_i-k_j},  \quad a_i=1 + i \frac{k_i}{\sigma} $$
and
 $$\eta_i=k_i X+(-k_i^3-3 k_i \sigma^2) T+\eta_i^0,  \quad i=1,2$$
 and $\xi_1$, $\xi_2$ are  arbitrary Grassmann-odd constants.

The novel feature of this solution compared to the ordinary Gardner is the {\it fermionic correction $\beta_{12}\xi_1\xi_2$}
 multiplying $A_{12}$ and the {\it fermionic dressing} of the parameter $\xi_i$ in the exponential by the factors $\alpha_{ij}$. 
Their role is more visible in the three-supersoliton solution, which have the form:

$$f=1 +a_1 e^{ \eta_1+\theta\xi_1}+a_2e^{\eta_2+\theta\xi_2}+ a_3e^{\eta_3+\theta\xi_3}+ A_{12} a_1 a_2(1+\beta_{12}\xi_1\xi_2)e^{ \eta_1 + \eta_2+\theta(\alpha_{12}\xi_1+\alpha_{21}\xi_2)}+$$
$$+ A_{13} a_1 a_3(1+\beta_{13}\xi_1\xi_3)e^{ \eta_1 + \eta_3+\theta(\alpha_{13}\xi_1+\alpha_{31}\xi_3)}+ A_{23} a_2 a_3(1+\beta_{23}\xi_2\xi_3)e^{ \eta_2 + \eta_3+\theta(\alpha_{23}\xi_2+\alpha_{32}\xi_3)}+$$
$$+  A_{12}  A_{13} A_{23} a_1 a_2 a_3 (1+\beta_{12}\alpha_{13}\alpha_{23}\xi_1\xi_2)(1+\beta_{13}\alpha_{12}\alpha_{32}\xi_1\xi_3)(1+\beta_{23}\alpha_{21}\alpha_{31}\xi_2\xi_3) \times$$
$$\times e^{\eta_1+ \eta_2 + \eta_3+\theta(\alpha_{12}\alpha_{13}\xi_1+\alpha_{21}\alpha_{23}\xi_2+\alpha_{31}\alpha_{32}\xi_3)},$$

$$g=1 +a^*_1 e^{ \eta_1+\theta\xi_1}+a^*_2e^{\eta_2+\theta\xi_2}+ a^*_3e^{\eta_3+\theta\xi_3}+ A_{12} a^*_1 a^*_2(1+\beta_{12}\xi_1\xi_2)e^{ \eta_1 + \eta_2+\theta(\alpha_{12}\xi_1+\alpha_{21}\xi_2)}+$$
$$+ A_{13} a^*_1 a^*_3(1+\beta_{13}\xi_1\xi_3)e^{ \eta_1 + \eta_3+\theta(\alpha_{13}\xi_1+\alpha_{31}\xi_3)}+ A_{23} a^*_2 a^*_3(1+\beta_{23}\xi_2\xi_3]e^{ \eta_2 + \eta_3+\theta(\alpha_{23}\xi_2+\alpha_{32}\xi_3)}+$$
$$+  A_{12}  A_{13} A_{23} a^*_1 a^*_2 a^*_3 (1+\beta_{12}\alpha_{13}\alpha_{23}\xi_1\xi_2)(1+\beta_{13}\alpha_{12}\alpha_{32}\xi_1\xi_3)(1+\beta_{23}\alpha_{21}\alpha_{31}\xi_2\xi_3) \times$$
$$\times e^{\eta_1+ \eta_2 + \eta_3+\theta(\alpha_{12}\alpha_{13}\xi_1+\alpha_{21}\alpha_{23}\xi_2+\alpha_{31}\alpha_{32}\xi_3)},$$
where $A_{ij}, \beta_{ij}, \alpha_{ij}, a_i, \eta_i$ cu $i=1,2,3$ are defined above and $\xi_1$, $\xi_2$ and $\xi_3$ are  arbitrary Grassmann-odd constants.
In the three-supersoliton solution, the fermionic corrections are also dressed because any pair of solitons interacts with a third supersoliton. This is clear from the expression:
$$A_{12}  A_{13} A_{23} a_1 a_2 a_3 (1+\beta_{12}\alpha_{13}\alpha_{23}\xi_1\xi_2)(1+\beta_{13}\alpha_{12}\alpha_{32}\xi_1\xi_3)(1+\beta_{23}\alpha_{21}\alpha_{31}\xi_2\xi_3)$$
$$\times e^{\eta_1+ \eta_2 + \eta_3+\theta(\alpha_{12}\alpha_{13}\xi_1+\alpha_{21}\alpha_{23}\xi_2+\alpha_{31}\alpha_{32}\xi_3)}.$$

Specifically, the fermionic correction in the case of a four-supersoliton solution the fermionic correction is
$\beta_{12}\alpha_{13}\alpha_{14}\alpha_{23}\alpha_{24}\xi_1\xi_2,$
because the pair of solitons 1 and 2 are interacting with soliton 3 and 4. Also, the dressing in the parameter $\xi_1$ in the exponent is $\alpha_{12}\alpha_{13}\alpha_{14}\xi_1$ because soliton 1 interacts with the three other solitons 2,3 and 4, and so on.

It was noted that in the case of integrable equations, interactions of more than two solitons are expressed as products of "two-body" solitons. Hence, if the procedure for interaction of three solitons is known, then it can be generalized to the case of $N$-solitons, thus proving the so-called complete Hirota integrability. 
In the supersymmetric context, because we have found the three-supersoliton solution, it is likely that the N-supersolitons have the form:
\vspace{1 cm}
$$f=\sum_{\mu \in \{0,1\}}\exp\left[\sum_{i=1}^N\mu_i\left(\eta_i+\log a_i+\theta \xi_i \prod_{m \neq i} \alpha_{im}\right)+\right.$$
$$\left.+\sum_{i<j}\mu_i\mu_j\left(\log A_{ij}+\beta_{ij} \xi_i \xi_j \prod_{k \neq {i,j}} \alpha_{ik}\alpha_{jk}\right)\right],$$
$$g=\sum_{\mu \in \{0,1\}}\exp\left[\sum_{i=1}^N\mu_i\left(\eta_i+\log a_i^*+\theta \xi_i \prod_{m \neq i} \alpha_{im}\right)+\right.$$
$$\left.+\sum_{i<j}\mu_i\mu_j\left(\log A_{ij}+\beta_{ij} \xi_i \xi_j \prod_{k \neq {i,j}} \alpha_{ik}\alpha_{jk}\right)\right].$$
where the products on $\alpha_{im}$ and $\alpha_{ik}\alpha_{jk}$ are respectively assumed to be unity for $N=1$ and $N=2$. 

Therefore, the super-Gardner equation is completely integrable.

\subsection{Super-rational and mixed super-rational supersoliton solutions}

It is well known that the ordinary Gardner equation also has rational nonsingular solutions of a Lorentzian shape and that they can be obtained using a limit procedure developed long ago by Ablowitz and Satsuma \cite{ablowitz}. 

In general, to obtain a $N$-rational solution from the $N$-soliton solutions, we must:

1. expand the tau functions corresponding to $N$-soliton solution in power series of $k_i$ ($k_i$ are the soliton parmeters),

2. choose the arbitrary phase constants ($\eta_i^0$) such that all terms of order $O(k_j^m), m<N$, are zero for all $j$;

3. send $k_j\to 0$ for all $j$, and obtain the $N$-rational solution from the remaining terms of order $O(k_j^p), p\geq N$.

We want to see if the focusing super-Gardner has nonsingular super-rational solutions. We use the above  method. 

Starting with the one-supersoliton solution, we set $k=\epsilon k_0$ and  $\xi=\epsilon \xi_0$, where $\epsilon \rightarrow 0$ 
(for two supersoliton parameters, we ensure that limit of both $k_0$ and $\xi_0$ are $O(1)$ terms). Choosing $e^{\eta^0}=-1$, we write the one-superrational solution as:
$$\Phi=D\phi=i D\log\frac{-\frac{i k_0}{\sigma}+ k_0  X -3 k_0 \sigma^2 T+  \theta \xi_0}{\frac{i k_0}{\sigma} + k_0  X  -3 k_0 \sigma^2 T+  \theta \xi_0}=$$
$$=-\frac{\sigma}{k_0}\frac{1}{\sigma^2 (X-3 \sigma^2 T)^2+1}\xi_0-2 \sigma\frac{1}{\sigma^2 (X-3 \sigma^2 T)^2+1}\theta. $$

Hence, we obtained both bosonic and fermionic components of a weakly localized solitary wave with a Lorentzian shape (lump) propagating with a fixed velocity $\sigma$. 
We can find the two-, three-, ... superrational solution similarly, but having the same speed, they do not interact. It is more interesting to see the interaction between a superrational solution and a supersoliton. For this, we apply the limit procedure to the two-supersoliton solution, but only on the first supersoliton (characterised by $(k_1,\xi_1)$) leaving the second supersoliton unchanged. As a result, setting  $k_1=\epsilon k_{10}$ and  $\xi_1=\epsilon \xi_{10}$, where $\epsilon \rightarrow 0$, and again choosing $e^{\eta_1^0}=-1$ in (\ref{2 ss Gardner}), we obtain:
$$f=Q+a_2 \left(\frac{4 k_{10}}{k_2}+Q\right)e^{\eta_2}+\frac{2}{k_2}a_2e^{\eta_2} \xi_{10} \xi_2+\theta\left(\left(a_2e^{\eta_2}-1\right)\xi_{10}+a_2  \left(\frac{2 k_{10}}{k_2}+Q\right) e^{\eta_2}\xi_2\right)$$
$$g=Q^*+a_2^* \left(\frac{4 k_{10}}{k_2}+Q^*\right)e^{\eta_2}+\frac{2}{k_2}a_2^*e^{\eta_2} \xi_{10} \xi_2+\theta\left(\left(a_2^*e^{\eta_2}-1\right)\xi_{10}+a_2^*  \left(\frac{2 k_{10}}{k_2}+Q\right)e^{\eta_2}\xi_2\right)$$
where:
$$Q=-k_{10}X+3 k_{10} \sigma^2 T-\frac{i k_{10}}{\sigma}, \qquad a_2=1+\frac{i k_2}{\sigma}.$$

To see the interaction effect, we fix the supersoliton reference frame ($\eta_2$=const) and obtain
\begin{equation*}
\lim_{t\to\pm\infty} \Phi_{mixed}=iD \log \frac{1 + a^*_2e^{\eta_2 + \theta\xi_2}}{1 + a_2 e^{\eta_2 + \theta\xi_2}},
\end{equation*}
which means that the superrational solution, as expected, does not change the supersoliton structure after the interaction.

On the other hand, if we fix the superrational solution frame ($Q=$const), then we obtain
\begin{equation}
\lim_{t\to+\infty} \Phi_{mixed}=iD \log \frac{Q^* - \theta\xi_{10}}{Q - \theta\xi_{10}},
\end{equation}
\begin{equation*}\label{limita}
\lim_{t\to-\infty} \Phi_{mixed}=iD\log\frac{Q^*-\theta\xi_{10}+4k_{10}/k_2+2\theta(k_2\xi_{10}-k_{10}\xi_2)/k_2+2\xi_{10}\xi_2/k_2}
{Q-\theta\xi_{10}+4k_{10}/k_2+2\theta(k_2\xi_{10}-k_{10}\xi_2)/k_2+2\xi_{10}\xi_2/k_2}.
\end{equation*}
Hence, the superrational solution is modified by the term $4k_{10}/k_2+2\theta(k_2\xi_{10}-k_{10}\xi_2)/k_2+2\xi_{10}\xi_2/k_2$. 
It is interesting that 
we {\it cannot} gauge away $\xi_0$ or $\xi_{10}$ and, in addition, the interaction has the same form as in the supersolitonic case \cite{fane1}.

\section{Defocusing super-Gardner equation}
We consider the equation
\begin{equation}\label{defo s-gard}
\Phi_t+\Phi_{xxx}-3(D\Phi)(\Phi D\Phi)_x-3\sigma(\Phi D\Phi)_x=0,
\end{equation}
where $\Phi(x,t,\theta)=\zeta(x,t)+\theta v(x,t)$ is a fermionic superfield. 

Setting $\Phi=D\log(g/f)$ and applying the Hirota bilinear formalism, we obtain

\begin{eqnarray}\label{bilinear}
({\bf D}_T+{\bf D}_X^3+3\sigma{\bf D}_X^2)g\cdot f=0\\
({\bf S}_X^3 +\sigma{\bf S}_X)g\cdot f=0, \nonumber
\end{eqnarray}
where $X=x-3 \sigma^2 t$ and $T= t$.

The structure of the supersoliton solutions of (\ref{defo s-gard}) is the same as that obtained in the focusing case. There are some essential differences that are noted in what follows.

The one-supersoliton solution is written as
\begin{eqnarray}\label{1sol  defo}
f&=&1+a e^{\eta+\theta\xi},\\
g&=&1+b e^{\eta+\theta\xi},\nonumber
\end{eqnarray}
where $\eta=k X+\omega T$, $\omega=-k^3+3 k \sigma^2$, $a=1 + \frac{k}{\sigma}$, $b=1 - \frac{k}{\sigma}$, and $\xi$ is an arbitrary Grassmann-odd constant. We have
\begin{eqnarray}\label{1 ss Gardner}
\Phi_{1ss}&=&D\phi=D \log \frac{g}{f}=D \log \frac{1+b e^{\eta+\theta\xi}}{1+ae^{\eta+\theta\xi}}=\\
&=&-\frac{k}{\sigma}\frac{\xi}{1+\sqrt{a b}\cosh \left(\eta+\log{\sqrt{a b}}\right)}- \frac{k^2}{\sigma}\frac{\theta}{1+\sqrt{a b}\cosh \left(\eta+\log{\sqrt{a b}}\right)}. \nonumber
\end{eqnarray}
For $\sigma=-2$ the bosonic part of our solution is the same as the one obtained by Chen and Liu in \cite{Chen}. 

The two-supersoliton solution is
\begin{eqnarray*}\label{2 ss defo Gardner}
f=1 +a_1 e^{ \eta_1+\theta\xi_1}+a_2e^{\eta_2+\theta\xi_2}+ A_{12} a_1 a_2(1+\beta_{12}\xi_1\xi_2)e^{ \eta_1 + \eta_2+\theta(\alpha_{12}\xi_1+\alpha_{21}\xi_2)},\\ \nonumber
g=1 +b_1 e^{ \eta_1+\theta\xi_1}+b_2e^{\eta_2+\theta\xi_2}+ A_{12} b_1 b_2(1+\beta_{12}\xi_1\xi_2)e^{ \eta_1 + \eta_2+\theta(\alpha_{12}\xi_1+\alpha_{21}\xi_2)},
\end{eqnarray*}
where
$$A_{ij}=\left(\frac{k_i-k_j}{k_i+k_j}\right)^2,  \quad \beta_{ij}=\frac{2}{k_i-k_j},$$  $$\alpha_{ij}=-\alpha_{ji}=\frac{k_i+k_j}{k_i-k_j},  \quad a_i=1 + \frac{k_i}{\sigma}, \quad b_i=1 - \frac{k_i}{\sigma} $$
and
 $$\eta_i=k_i X+(-k_i^3+3 k_i \sigma^2) T,  \quad i=1,2,$$
 and $\xi_1$ and $\xi_2$ are  arbitrary Grassmann-odd constants.

\vspace{0.4 cm}

{\bf Super-shock waves and mixed solutions.}

Unlike focusing super-Gardner, the defocusing case admits shock wave solutions. They appear only for $k=-\sigma$. Introducing this condition  in (\ref{1sol  defo}), we obtain:
\begin{eqnarray*}\label{shock wave}
\Phi=D\phi=D\log \frac{g}{f}&=& \frac{1}{2}\left(1+\tanh \left(-\frac{\sigma}{2}(X+2 \sigma^2 T)+\frac{\log 2}{2}\right)\right)\xi+ \nonumber \\ 
&- &\frac{\sigma}{2}\left(1+\tanh \left(-\frac{\sigma}{2}(X+2 \sigma^2 T)+\frac{\log 2}{2}\right)\right) \theta.
\end{eqnarray*}

The bosonic part of this solution with $\sigma=-2$ coincides with the shock wave solution of the defocusing Gardner equation obtained in \cite{Chen}.

We can also have a mixed solution (one supersoliton and one supershock wave):

\begin{eqnarray}\label{mixed}
\Phi=D\phi=D\log \frac{
1 +b_1 e^{ \eta_1+\theta\xi_1}+2e^{\eta_2+\theta\xi_2}+2 A_{12} b_1(1+\beta_{12}\xi_1\xi_2)e^{ \eta_1 + \eta_2+\theta(\alpha_{12}\xi_1+\alpha_{21}\xi_2)}}
{1 +a_1 e^{ \eta_1+\theta\xi_1}}, \nonumber
\end{eqnarray}
where $\beta_{12},b_1,A_{12},a_1$ are the same as above.

\section{Conclusions}
We have used the superbilinear formalism to analyse the integrability and supersoliton solutions of a supersymmetric extension of the Gardner equation. In contrast to the classical case, this extension cannot be obtained by imposing a nonzero boundary condition. We proved that the equation admits at least three-supersoliton solution and is hence completely integrable. Because the supersoliton solutions are rich, we also discussed superrational solutions obtained by taking the so-called long-wave limit developed long ago by Ablowitz and Satsuma. We showed that we have the same behaviour in the interaction between the superrational and supersoliton solutions as in the classical case: the supersoliton solution is unchanged, whilst the superrational solution aquires a $\theta$-dependent shift. We also analyzed the case of the defocusing supersymmetric Gardner equation and constructed the supersoliton and super-shock wave solutions.  
\vskip 0.5cm
{\bf Acknowledgements:} NBC is supported by the project PN-II-ID-PCE-2011-3-0083, Romanian Ministery of Education and Research. ASC is supported by the project PN-II-ID-PCE-2011-3-0137, 
Romanian Ministery of Education and Research.
\section{Appendix}
We list here several properties of the supersymmetric Hirota bilinear operator useful for deriving bilinear forms and supersoliton solutions:
\begin{eqnarray}
\mathbf{S}_x^{2N} a \cdot b&=&\mathbf{D}_x^N a \cdot b, \nonumber\\
\mathbf{S}_x^{2N+1} a\cdot b&=&\mathbf{S}_x\mathbf{D}_x^N a \cdot b, \nonumber\\
\mathbf{S}_x a\cdot b&=&(\mathcal{D}a)b-(-1)^{|a|}a (\mathcal{D}b),\nonumber\\
\mathbf{S}_x^{2N+1} e^{\eta_1} \cdot e^{\eta_2}&=&[
{\zeta_1}-
{\zeta_2}+\theta (k_1-k_2)](k_1-k_2)^N e^{\eta_1+\eta_2},\nonumber\\
\mathbf{S}_x^{2N+1} 1 \cdot e^{\eta}&=&(-1)^{N+1}(
{\zeta}+\theta k)k^N e^\eta=(-1)^{N+1}\mathbf{S}_x^{2N+1} e^{\eta} \cdot 1,\nonumber
\end{eqnarray}
where $\eta_i=k_ix+\theta {\zeta_i}$ and ${\zeta_i}$ are odd Grassmann numbers.

\end{document}